\documentclass[12pt,preprint]{aastex}

\newcommand{\unit}[1]{\,\mathrm{#1}}

\begin{document}
\title{Did the Milky Way dwarf satellites enter the halo as a group?}

\author{Manuel Metz, Pavel Kroupa}
\affil{Argelander-Institut f\"ur Astronomie, Universit\"at Bonn, Auf dem H\"ugel 71, D--53121 Bonn, Germany}
\email{mmetz,pavel@astro.uni-bonn.de}

\author{Christian Theis, Gerhard Hensler}
\affil{Institute of Astronomy, University of Vienna, T\"urkenschanzstrasse 17, 1180 Vienna, Austria}

\author{Helmut Jerjen}
\affil{Research School of Astronomy and Astrophysics, ANU, Mt. Stromlo Observatory, Weston ACT 2611, Australia}

\begin{abstract}
The dwarf satellite galaxies in the Local Group are generally considered to be hosted in dark matter subhalos that survived the disruptive processes during infall onto their host halos. It has recently been argued that if the majority of satellites entered the Milky Way halo in a group rather than individually, this could explain the spatial and dynamical peculiarities of its satellite distribution. Such groups were identified as dwarf galaxy associations that are found in the nearby Universe. In this paper we address the question whether galaxies in such associations can be the progenitors of the Milky Way satellite galaxies. We find that the dwarf associations are much more extended than would be required to explain the disk-like distribution of the Milky Way and Andromeda satellite galaxies. We further identify a possible minor filamentary structure, perpendicular to the supergalactic plane, in which the dwarf associations are located, that might be related to the direction of infall of a progenitor galaxy of the Milky Way satellites, if they are of tidal origin.
\end{abstract}


\section{Introduction}\label{sec_intro}
Dwarf galaxies in the Local Group, and satellite galaxies of the Milky Way (MW) and Andromeda (M31) in particular, are of great importance to understand the physics of structure formation on galaxy scales because they can be studied in unsurpassable detail. Commonly they are associated with cosmological sub-structures that entered the Milky Way halo and it has been inferred, assuming that the dwarf galaxies are in virial equilibrium, that they are dominated by a massive dark matter component though the nature of this component is still disputed \citep{gilmo07}. There are, however, some puzzling findings not yet completely explained.

For instance, it had been noticed that the number of observed satellite galaxies in the Local Group is at least an order of magnitude smaller than expected from cold dark matter (CDM) simulations \citep{moore99, klypi99,diema08}. Different scenarios were subsequently proposed to explain the satellite galaxies in the Local Group in the context of being CDM dominated subhalos: \citet{stoeh02} argued that only the most massive subhalos were able to form stars and the rest thus remains dark. \citet{libes05} proposed that current satellite galaxies are those with the most massive progenitors, also found by \citet{strig07} who alternatively found accordance with the earliest forming halos. In contrast, \citet{sales07a} argued for low-mass systems being the origin of satellite galaxies, or it has been suggested that observers just overlooked all the hundreds of satellite galaxies surrounding the Milky Way due to their extreme low star densities \citep{tolle08}.

Recently the interesting scenario was put forward that the satellite galaxies did not enter the Milky Way halo individually in a random fashion, but rather in groups. \citet{li08} argued that if only one or two groups fell into the MW halo this could account for the observed highly anisotropic spatial appearance of MW companions, the disk of satellites \citep*[DoS,][]{metz07}. \citet{dongh08} suggested that groups formed within LMC-like dark matter host-halos and that these hosts are able to produce many more satellites in subhalos than in a  MW-like host-halo. They predicted that such groups are visible today and linked them as the dwarf associations found by \citet{tully06}.

Here we study the properties of the dwarf associations, and compare them to the satellite galaxies of the Milky Way. In Section~\ref{sec_mggroup} we discuss the membership of satellites in the proposed Magellanic Group. The implications of the extent of observed dwarf associations is addressed in Section~\ref{sec_extend} and their spatial locations are analysed in Section~\ref{sec_loc}. We finally discuss the proposed scenario in Section~\ref{sec_discussion} and summarize our conclusions in Section~\ref{sec_conclusions}.

\section{Associations of dwarf galaxies}\label{sec_dga}

\subsection{A Magellanic Group?}\label{sec_mggroup}
\citet{dongh08} proposed that the majority of the brighter satellite galaxies of the Milky Way originate from a single group of dwarf galaxies in dark matter subhalos with their main component being the LMC that fell late into the Milky Way halo. Indeed, such associations of dwarf galaxies exist within $8\unit{Mpc}$ of the Local Group \citep{tully06}.

Specifically, \citeauthor{dongh08} proposed that the Magellanic Clouds together with the Sagittarius, Ursa Minor, Draco, Sextans, and Leo~II dwarf galaxies once belonged to such a group (hereafter DL08 sample). This appears to be a rather ad-hoc selected sample and is based on a mix of different and disagreeing references. Earlier proposed streams of MW satellite galaxies, that are referred to, consist of different objects \citep[LMC, SMC, Dra, UMi, \& Scl;][]{lynde76}, and a second different plane as proposed by \citeauthor{kunke76} \citep[LMC, SMC, Dra, UMi, Car, Leo~I \& II;][]{kunke76,kunke79} is $\sim 40\degr$ off the one proposed by \citeauthor{lynde76}. Later, \citeauthor{lynde82a} associated Fornax, Leo~I, Leo~II, and Sculptor to the so-called FLS stream, whereas he confided Carina to the LMC group \citep{lynde82,lynde82a}. It is important to note that two of the galaxies in the DL08 sample, Sextans and Sagittarius, were not known until the 90s \citep{irwin90,ibata94}, and thereafter \citet*{kroup05} and \citet{metz07} did not identify an individual stream, but rather highlighted the fact that \emph{all} satellite galaxies of the MW belong to a virtual plane -- the disk of satellites.

In particular one finds that the Sagittarius dwarf galaxy today must be on a quite different orbit than the other galaxies in the proposed ensemble. Based on its location and proper motion measurements one arrives at an orbit that is perpendicular to that of the Magellanic Clouds \citep{palma02,metz08}. Furthermore, models describing the Sagittarius Stream suggest rather short orbital time-scales for Sagittarius, $t_{\rm orb} <1\unit{Gyr}$, small Galactic apocentric distances, $\sim 60\unit{kpc}$, and that the dwarf must have been on this close path for several orbits \citep{ibata97,law05,fellh06}. This can only be explained in the above scenario by an early scattering event with the LMC that brought Sagittarius into a close orbit \citep{zhao98, sales07b}. However, this is inconsistent with the recent measurements of the proper motions of the LMC/SMC system \citep{kalli06a,kalli06b,piate08}. These data suggest that the Magellanic clouds are on their first passage about the Milky Way or are on a very wide orbit \citep{besla07,wu08}. Consequently, the LMC is on a rather different orbit from that of Sgr.

Proper motion data are currently available for another seven Milky Way companion galaxies, including Fornax and Carina \citep{piate03,dines04,piate07}. Combining the data, \citet{metz08} find that these two are compatible with the hypothesis that they are in a similar orbital plane as the LMC/SMC, but both are missing in the DL08 sample. In summary, the selection of members of the LMC group as proposed by \citeauthor{dongh08} is not supported by the available data. This, however, does not invalidate the suggestion that the DoS is a result of group infall.

\subsection{The extent of dwarf associations}\label{sec_extend}
\begin{figure}
  \plotone{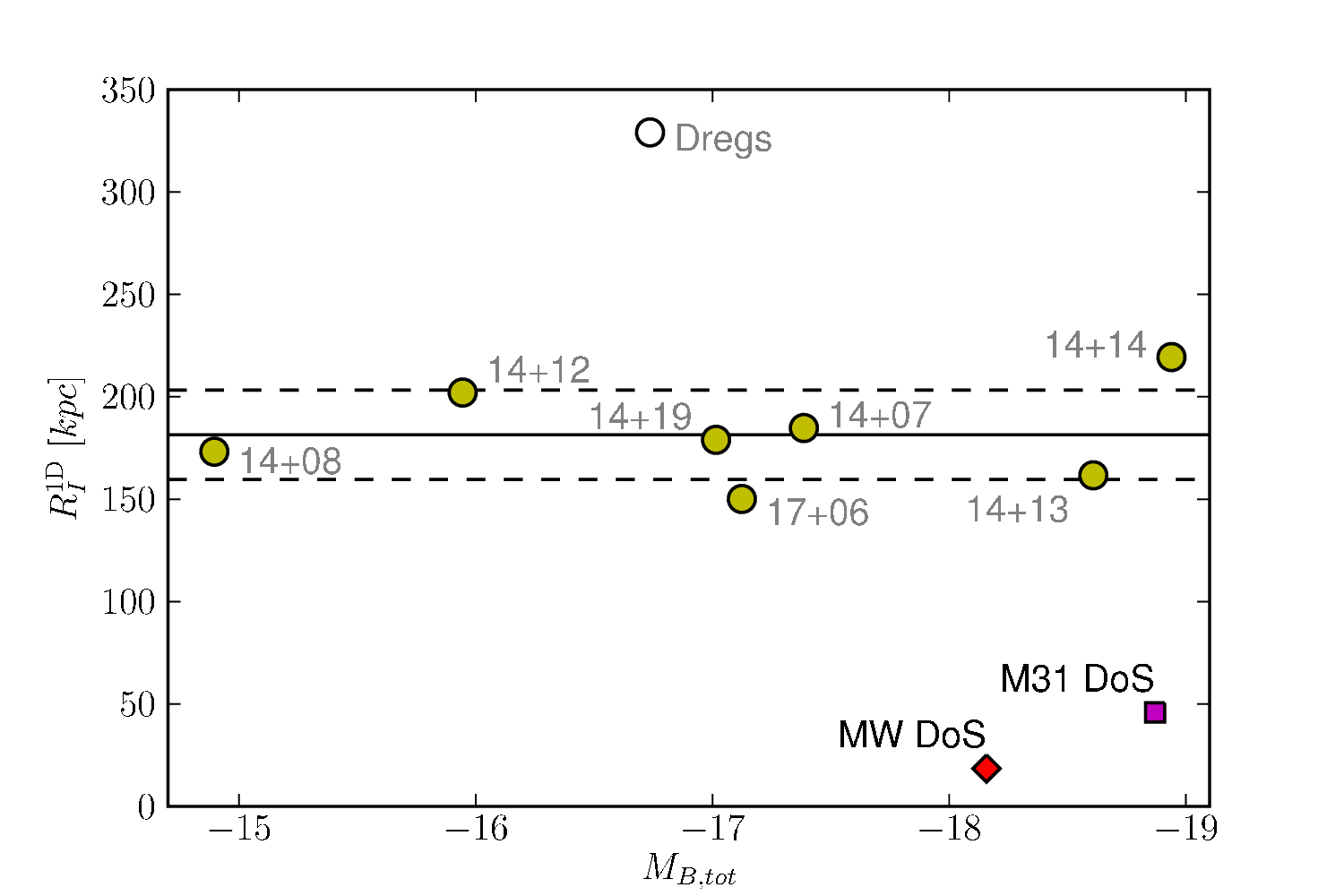}
  \caption{The characteristic one-dimensional size-scale of associations of dwarf galaxies, $R_I^{\rm 1D}$, plotted versus the total absolute B-band magnitude of the association. The horizontal lines give the mean and the standard deviation of $R_I^{\rm 1D}$ of the sample, excluding the Dregs with $R_I^{1\rm D}=329 \unit{kpc}$ which is shown by the open circle. The diamond symbol shows, for comparison, the height of the disk of satellites of the Milky Way versus the total of absolute B-band magnitude of the Magellanic Clouds \citep[taken from][]{karac04}, and the square is for the Andromeda system \citep{metz07}. No inertial radius is given for 14$-$14 which is a pair of galaxies.}
  \label{fig_RI1D}
\end{figure}

When an association of dwarf galaxies falls into a large host-halo, the group gets stretched and wound up on its orbit in the halo about the host galaxy \citep{li08}. The extent of the pre-infall association is characterized by $R_I^{\rm 3D} = \left( \sum_{i=1}^N r_i^2/N \right)^{1/2}$, which is the three-dimensional inertial radius of the association as given in \citet{tully06}. Here, $r_i$ is the distance of a galaxy from the geometrical center of a group and $N$ is the number of member galaxies in the group. After the breakup, a one-dimensional characteristic scale, the thickness of the remaining distribution of dwarf galaxies -- if interpreted as a flattened structure -- can be derived. It is expected that such a break-up leads to a flattened structure with mean height $R_I^{\rm 1D} = R_I^{\rm 3D} /\sqrt{3}$. Using the observed values for the nearby associations 14+12, 14+13 (excluding IC~5152), 14+07, 14+08, 17+06, 14+14, 14+19, and Dregs as given by \citet{tully06}, we find values ranging from $R_I^{\rm 1D} = 150 \unit{kpc}$ to $210\unit{kpc}$, and even up to $390\unit{kpc}$ if we include the loose group of the `Dregs'. These are plotted in Figure~\ref{fig_RI1D} against the integrated B-band luminosity of the groups. The solid horizontal line indicates the mean $\langle R_I^{\rm 1D} \rangle = 181\unit{kpc}$ (excluding the Dregs) with a standard deviation of $22\unit{kpc}$ shown by the dashed lines. Compared to the height $\Delta=18.5\unit{kpc}$ \citep{metz07} of the disk of the classical satellites of the Milky Way, shown by the diamond symbol in Fig.~\ref{fig_RI1D}, all these values are an order of magnitude (at least $4.5\sigma$) larger than those found for the MW. Also the height of the DoS of Andromeda, $\Delta=46\unit{kpc}$, is significantly smaller than what is found for the dwarf associations. To test the robustness of our result, we also re-calculated the one-dimensional characteristic scale for all the associations, by excluding their most distant group-member. In that case the smallest value derived is $R_1^{\rm 1D*}=92\unit{kpc}$, and the mean is $\langle R_1^{\rm 1D*} \rangle = 129 \pm 23 \unit{kpc}$, which is significantly larger than the height of the DoS. If we assume that the smaller RMS values are purely caused by the statistical effect of a larger sample size of the classical DoS, this can also not account for the difference: in this case we would expect at most a height of the order $25\unit{kpc}$ for the MW, still $\sim 4\sigma$ smaller than for the associations. \emph{If} the disk of satellites were to originate from the late infall of an association of dwarf galaxies similar to those observed near the Local Group today, it is incomprehensible how such a distribution tightens that much to become a structure similar to the DoS.

\bigskip
The size of the observed associations of dwarf galaxies has another very important consequence if one assumes that they are the prototypes of an LMC group aged in isolation as proposed by \citet{dongh08}. Their suggested model is based on the idea that gas is blown out of a subhalo (by whatever process) and thermalizes to the virial temperature of the host halo. So this model implicitly assumes that a small halo -- that will eventually become a dwarf satellite galaxy -- is embedded in a larger host-halo. Because the virial temperature of an LMC-like host-halo is lower than for a MW like galaxy, they argued that the cooling of the gas is more efficient in the smaller host-halo and the gas settles back into the embedded subhalos. Taking their assumed values, we find that the mass ratio of the host halos is of order $M_{\rm LMC}/M_{\rm MW} = 1/10$. This mass ratio is related to the ratio of the virial radii of the host halos by $R_{\rm vir, LMC} = \sqrt[3]{M_{\rm LMC}/M_{\rm MW}} \, R_{\rm vir, MW}$ \citep[cf.\ for example][]{saluc00,bullo01} which gives $R_{\rm vir, LMC} \sim 115\unit{kpc}$ for $R_{\rm vir, MW} \sim 250\unit{kpc}$. If we now interpret $R_I^{3D}$ as a characteristic size-scale for the mutual distance between any two galaxies in a dwarf association, we see that $\langle R_I^{\rm 3D}\rangle = 314\unit{kpc}$ for the dwarf galaxy associations is almost a factor 3 larger than the assumed virial radius of the LMC. Thus it is immediately clear that dwarf galaxies are not deeply embedded in the host halo of the main component of the groups \citep[see also][]{tully06} and do not meet the assumptions made by \citeauthor{dongh08}.

It follows that today's dwarf galaxy associations cannot be the evolved prototypes of a dwarf galaxy group that fell into the MW halo, but that such a hypothetical configuration must then have been much more compact and does not correspond to any known existing structures.

\subsection{Locations of dwarf associations}\label{sec_loc}
\begin{figure}
  \plotone{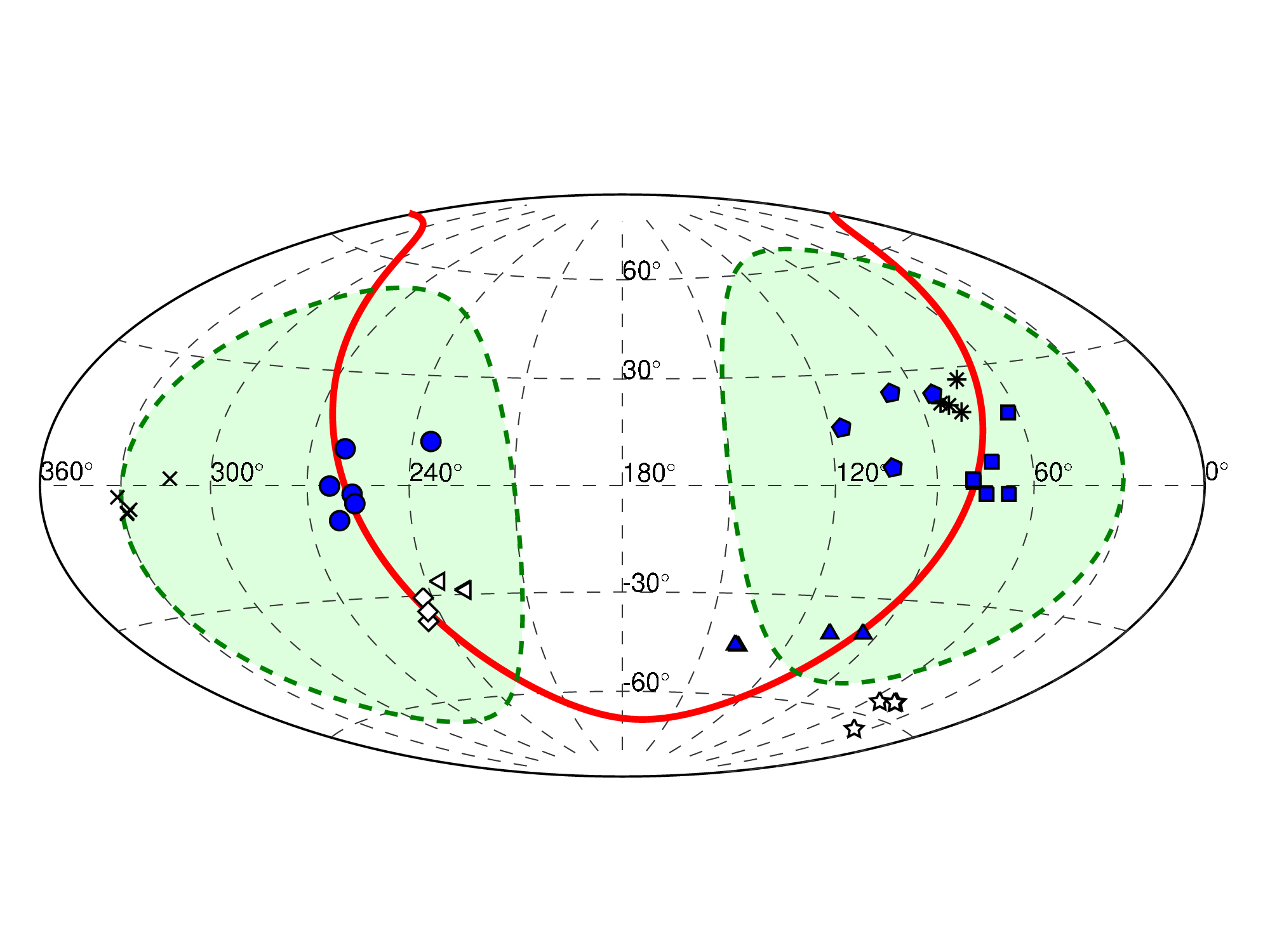}
  \caption{Projected locations of the associations of dwarf galaxies as given in \citet{tully06}, plotted in supergalactic coordinates with the supergalactic plane being the equator. Dwarf galaxies belonging to the same group are marked by corresponding symbols: filled triangles show the 14+12 Group, filled circles the 14+13 Group, filled squares the 14+7 Group, asterisks the 14+8 Group, crosses the 17+06 Group, open triangles the 14$-$14 Group, open diamonds the 14+14 Group, open stars the 14+19 Group, and filled pentagons the Dregs. The thick line shows the projected orientation of the disk of satellites of the Milky Way. The light shaded areas show regions on the sky with galactic latitude $|b|>30\degr$ that were searched for dwarf galaxy associations.}
  \label{fig_lgroups}
\end{figure}

One interesting property of the Milky Way satellite galaxies still remains to be examined in the context of the group infall scenario which has not been discussed in detail before: the orientation of the prolate halo in galaxy clusters is governed by the direction of the accretion along filamentary structures. Downscaling this behavior to a Milky Way-like galaxy, one expects that the spatial distribution of the satellite galaxies is correlated with the medium scale filamentary structure \citep{knebe04,baili05,libes07,li08}. It has, however, been shown \citep{metz07} that the disk of satellites of the MW is highly inclined with respect to the supergalactic plane \citep[SGP,][]{devau91} with an inclination angle of $\sim 69\degr$, i.e.\ almost perpendicular with respect to the medium-scale matter distribution, as is also indicated in Figure~\ref{fig_lgroups} by the thick line. If \emph{(i)} structures were accreted along this main medium-scale matter distribution and \emph{(ii)} these structures build up the disk of satellites this would not explain the orientation of the DoS.

In Figure~\ref{fig_lgroups} we show the projected locations of the associations of dwarf galaxies as identified by \citet[compare to their figure~13]{tully06} in the supergalactic coordinate system. For reference, we give here the orientation of the normal of the disk of satellites of the Milky Way in supergalactic coordinates: $(SGL, SGB)=(348.8\degr, -20.7\degr)$. Five out of the nine dwarf galaxy associations are found in projection close to the supergalactic plane, $|SGB| < 30\degr$, and thus are likely belonging to this structure. Three out of these associations, 14+13, 14+07, and 14+08 are, in projection, close to the DoS too. Interestingly, three out of the four remaining associations, 14+12, 14+14, and 14$-$14, are also located very close to the line that indicates the orientation of the disk of satellites. These associations are found at distances of $1.4$, $5.8$, and $4.4\unit{Mpc}$ from the Sun, respectively, but they seem \emph{not} to belong to the supergalactic plane.

\begin{figure}
  \plotone{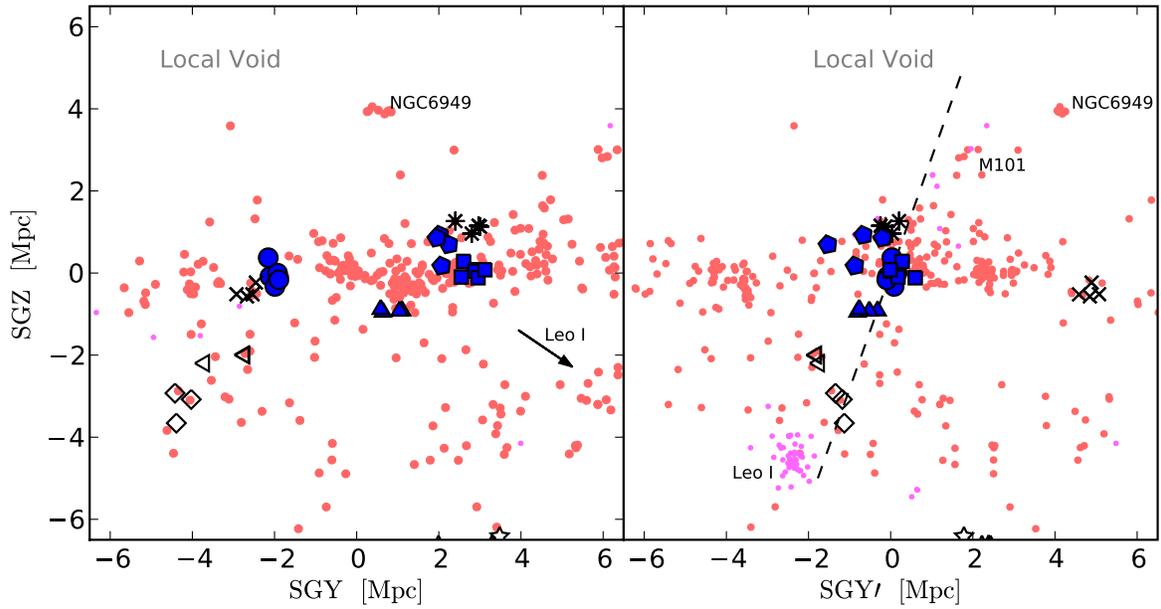}
  \caption{Locations of dwarf galaxies in associations, shown by the same markers as in Figure~\ref{fig_lgroups}, in supergalactic Cartesian coordinates. The right panel shows a view rotated about the SGZ-axis such that the DoS of the Milky Way is seen edge-on as indicated by the dashed line. The locations of nearby galaxies as listed in \citet{karac04} are shown by dots whereby smaller dots mark galaxies more distant than $8\unit{Mpc}$. Some galaxy groups are indicated in the panels; the arrow in the left panels shows the direction towards the Leo~I group.}
  \label{fig_sgcoords}
\end{figure}

For a better visualization of the real spatial distribution we plot the locations of the dwarf associations in Figure~\ref{fig_sgcoords} in supergalactic Cartesian coordinates. In the left panel supergalactic coordinates $SGY$ and $SGZ$ are shown. In the right panel, the distribution is rotated about the $SGZ$-axis such that the DoS of the Milky Way, indicated by the dotted line, is seen edge-on. It is evident from that plot that 14+12 and 14+14, marked by the filled triangles and open diamond symbols, indeed lie on the virtual extension of the disk of satellites of the Milky Way. We emphasize that the DoS is derived by finding the solution that minimizes the orthogonal distances to that plane of the MW satellite galaxies within only $\sim 250\unit{kpc}$, while the two associations are located at distances of $1\,400$ and $5\,800 \unit{kpc}$.

Moreover the locations of nearby galaxies within about $10\unit{Mpc}$ radius as listed in \citet{karac04} are marked in Figure~\ref{fig_sgcoords} by dots, smaller ones marking those galaxies at distances $>8\unit{Mpc}$. The extent of nearby galaxies in the SG plane of $\sim 2\unit{Mpc}$ is clearly visible in this plot. The 14+12 group is found at the periphery of the SG plane. While hardly any galaxies are detected towards the supergalactic north-pole region in the Local Void \citep[LV,][]{tully87} there is a crowding of galaxies in the opposite direction with no apparent structure. In the edge-on projection (right panel) the virtual extension of the DoS again is close to some prominent groups: one is the Leo~I group, but it is located at a distance of $~10.5\unit{Mpc}$. Another one is the M101 group at the edge of the Local Void. Offset from the plane, but also at the border of the LV, is the NGC6949 group.

The surprising finding that the dwarf associations, that are out of the supergalactic plane, are found close to the extension of the virtual plane of the MW DoS might have different explanations: the simplest one being that it is just a chance finding. As \citet{tully06} noted, large parts of the sky, those with galactic latitude $|b| \lesssim 30\degr$, i.e.\ about half of the total sky, have not been included in their search for dwarf galaxy associations. Two of the dwarf associations away from the supergalactic plane are found close to the DoS, but if another dozen associations are still awaiting discovery they could be way off the extended virtual plane of the MW DoS.
A second possible explanation is that the locations of the groups reflect a minor medium-scale structure, a less pronounced filament. If this were the case, it would mean that groups can also fall into the Milky Way host halo along this secondary preferred structure, which is compatible with the scenario as proposed by \citet{dongh08}.

\section{Discussion}\label{sec_discussion}
\citet{dongh08} did not characterize the properties of the group before infall studied in their simulation in a way that can be compared to the observed dwarf associations identified by \citet{tully06}. Here we showed that a prerequisite for the proposed scenario is that the subhalos are deeply embedded in the host halo of the parent galaxy. Such groups, if they would evolve in isolation, \emph{do not} resemble the dwarf galaxy associations we see today as has been conjectured by \citeauthor{dongh08}, because they would be much too compact in comparison to the observed associations. 

In contrast, such systems are likely similar to the groups as studied by \citet{li08}. Here groups were identified using a relative distance cut-off, $d < 40\unit{kpc}$, between pairs of subhalos at the time of accretion. So, these groups were already compact before the time of accretion, much more compact than today's dwarf galaxy associations. There is, however, an unanswered problem with the \citeauthor{li08} scenario, too: these authors studied \emph{one} MW-like halo in a CDM simulation. In that simulation they found that about one-third of the subhalos in total fell into the halo in groups. If a disk of satellites originates from the infall of order one group hosting luminous dwarf galaxies, this means that the majority of the groups have to remain dark, as well as most subhalos that individually entered the MW halo which are two-thirds of all subhalos in total. It is not evident which process should separate the two categories of subhalo groups, luminous and dark ones. If in contrast all of the groups host luminous dwarf galaxies the DoS can not be explained since the groups can come from any direction. Alternatively to explain the DoS in that scenario, most groups have to be very efficiently destroyed after they entered the host halo and only a few remain intact.

Additionally, the process of gas accretion into sub-halos embedded in LMC-type halos is not so obvious. Due to efficient cooling at the virial temperature of LMC-type halos \citet{dongh08} suggest that ram pressure stripping might be less efficient for embedded subhalos. They argue that these subhalos might keep more pristine gas and that they might even accrete cold (or cooled) gas, by this becoming luminous. However, details strongly depend on the inherent physical processes. In contrast, \citet{mayer07} found an early gas stripping in their simulations of dwarf satellite systems. Gas might for example be blown out by supernova activity in shallow subhalo potentials. Furthermore, the accretion of cold gas (e.g.\ clumps or filaments) is only feasible if the relative velocity falls below the virial velocity. A substantial ingredient to decide on the validity of the contrasting models could be provided by studies of particular element abundances. It has also been attempted to understand the formation of the galactic stellar halo by the accretion of dSphs but until now with controversial issues. Thus, the exact interplay of heating, cooling, gas dynamics, accretion, and star formation in a complex dark matter environment still has to be investigated in detail by means of numerical models.

It might be argued that compact groups of dwarf galaxies, if they existed at early times, are not observed today because if they had evolved in isolation over a Hubble time, all its members have already merged and thus we do not observe a group but only a remnant galaxy. The merger product observable today is likely a small early type galaxy \citep{naab03}. Only dwarf ellipticals found in isolation can potentially be directly linked with evolved compact groups of dwarf galaxies, because ellipticals found in a denser region can not be distinguished from ellipticals formed in a different manner. We scanned for nearby isolated dwarf ellipticals in the catalog of \citet{karac04}, and defined isolated dwarf ellipticals to have no other galaxy within a radius of $200\unit{kpc}$ and to be fainter in the B-band than the brightest dwarf galaxy associations, i.e.\ fainter than $M_B=-19$, because we assumed that after merging the potential remnant evolves passively, thus becoming fainter in the blue. Only one such elliptical could be identified in the catalog, NGC~855 at a distance of $9.7\unit{Mpc}$, having no other galaxy listed in the catalog within a radius of $1.1\unit{Mpc}$. \citet{walsh90} found extended HI emission in that dwarf elliptical and \citet{li07} suggested that it might have undergone a recent minor merger event. This would be consistent with the hypothesis that it is the merger remnant of a compact group of dwarfs. The problem with this particular candidate is that it is at the very periphery of the volume covered by the catalog and in a region likely not completely sampled. Thus it is unclear whether NGC~855 is really that isolated. Such objects are, however in any case, very rare, such that the compact-group infall scenario of \citet{li08} is unlikely able to explain that both, the MW and M31, have DoSs with comparable thickness.

There is a possibility that the central galaxy of the dwarf associations is surrounded by an additional number of dwarf spheroidals that have a too low luminosity to be included in the sample. The faintest galaxy in the associations identified by \citet{tully06} has an absolute magnitude of $M_B = -9.3$ (excluding the Dregs association). About half of the nine classical MW dSphs are fainter than this value (Leo~II has approximately the same absolute magnitude, $M_B=-9.2$). If we adopt an apparent magnitude limit of $B=17.5\unit{mag}$ for the detection of dwarfs \citep{karac04}, we find that two out of nine MW dSphs would be detected from a distance of $6\unit{Mpc}$. Six associations have distances $<6 \unit{Mpc}$ and four out of these are at distances $\lesssim 3\unit{Mpc}$. It appears likely that if an additional population of dwarfs exists that has a luminosity function similar to the one of the MW dSphs, at least in one of the associations one such dwarf would already have been found, even if they were surrounded by four or five such dwarfs only -- but to our knowledge no such companion has been found. This argument does of course not disprove the existence of a population of fainter satellites surrounding the main component (or all the dwarf galaxies) of the associations, but if it exists it appears likely that they must all be significantly fainter than $M_B=-9.3$, possibly more similar to the objects identified in the SDSS in recent years \citep{metz09} -- a scenario that can be tested observationally.

\section{Conclusions}\label{sec_conclusions}
We argue that there are two main issues of the proposed group infall model: first, no groups of dwarf galaxies are observed in the vicinity of the Local Group today that are consistent with the prediction by \citet{dongh08}, i.e.\ which are compact enough to produce a disk of satellites. One would need to postulate that such compact groups were common about 10 Gyr ago, and have disappeared by now. In this case there ought to be merger remnants of such groups of dwarfs. There are no good candidates for merger remnant objects, except possibly the dwarf elliptical NGC~855 which needs further investigation to confirm its properties and possible isolated status. Furthermore, it needs to be studied what properties groups of dwarf galaxies have if they evolve in isolation, whether they merge and how intermediate structures, groups that have not yet fully merged, look like. In any case, putative compact groups appear to have been very rare making the group infall scenario as an explanation for the disks of satellites of the MW and M31 very unlikely.

The second problem is that the group infall scenario requires that not only most of the subhalos are dark, but also that most of the groups are dark. Otherwise the disk of satellites can not be explained in that model, because this is only reproduced if on the order of one group fell into each, the MW and M31 halos.

An alternative scenario for the origin of the dwarf spheroidal satellite galaxies of the Milky Way goes back to an observation already made by \citeauthor{zwick56} in the 1950s. He pointed out that galaxies may form anti-hierarchically in the material thrown out off interacting large galaxies. These nowadays called tidal dwarf galaxies \citep[TDG,][]{mirab92} are well known to form in the Universe \citep[e.g.][]{weilb03,walte06}. 
Based on the identification of streams of satellite galaxies and globular clusters on the sky, in a pioneering work \citet{lynde83} had suggested that the dwarf spheroidals originate from the break up of a former larger galaxy. With the growing success of the dark matter theory this idea was, however, often overlooked and dwarf galaxies were naturally identified with accreted cosmological sub-structures. \citet{kroup05} highlighted the issue again deeming the great plane of the MW satellites to be inconsistent with the CDM theory. Indeed, there is a strong correlation of the orbital poles of the satellites \citep{palma02,metz08} suggesting a common origin. If the satellites were of tidal origin this would naturally account for the orbital correlation. A tidal origin is also consistent with the possible existence of a minor filamentary structure as shown in \S\ref{sec_loc}. If the progenitor of the satellite galaxies came along this direction the orientation of the DoS would be related to the infall direction as is the case for infalling dark matter halos along filaments. The major difference is that subhalos can come at any angle along the filament and thus in general do not have correlated orbits, whereas tidal dwarfs do.

In \emph{summary}, this paper presents a detailed investigation of recent claims that the dSph satellites of the Milky Way entered the Milky Way system as part of a galaxy group surrounding the Magellanic Clouds, or as a stand-alone dwarf group. By considering the observed properties of the known groups of galaxies within about 6 Mpc of the Milky Way it is concluded that all these systems are too spatially extended to be the progenitors of the disk-like Milky Way satellite distribution. Because the observed galaxy groups are not compact, the recent claims that dwarf galaxies within such groups would be able to retain, or even re-accrete, gas are rendered unlikely. Tentative evidence for an intermediate-scale filamentary structure in the distribution of nearby galaxy groups is brought up which could indicate the direction of infall of a galaxy which interacted with the young Milky Way such that tidal dwarf galaxies formed in the expanding tidal arms producing a system of dSph satellite galaxies that remain correlated in phase-space.

\acknowledgements
MM acknowledges support by the DFG Priority Program 1177. MM, PK, and HJ acknowledge the financial support of two DAAD/Go8 travel grants.

\bibliographystyle{apj}
\bibliography{quotes,quotes3,quotes2,quotes_preprint}

\begin{thebibliography}{54}
\expandafter\ifx\csname natexlab\endcsname\relax\def\natexlab#1{#1}\fi

\bibitem[{{Bailin} \& {Steinmetz}(2005)}]{baili05}
{Bailin}, J. \& {Steinmetz}, M. 2005, \apj, 627, 647

\bibitem[{{Besla} {et~al.}(2007){Besla}, {Kallivayalil}, {Hernquist},
  {Robertson}, {Cox}, {van der Marel}, \& {Alcock}}]{besla07}
{Besla}, G., {Kallivayalil}, N., {Hernquist}, L., {Robertson}, B., {Cox},
  T.~J., {van der Marel}, R.~P., \& {Alcock}, C. 2007, \apj, 668, 949

\bibitem[{{Bullock} {et~al.}(2001){Bullock}, {Kolatt}, {Sigad}, {Somerville},
  {Kravtsov}, {Klypin}, {Primack}, \& {Dekel}}]{bullo01}
{Bullock}, J.~S., {Kolatt}, T.~S., {Sigad}, Y., {Somerville}, R.~S.,
  {Kravtsov}, A.~V., {Klypin}, A.~A., {Primack}, J.~R., \& {Dekel}, A. 2001,
  \mnras, 321, 559

\bibitem[{{de Vaucouleurs} {et~al.}(1991){de Vaucouleurs}, {de Vaucouleurs},
  {Corwin}, {Buta}, {Paturel}, \& {Fouque}}]{devau91}
{de Vaucouleurs}, G., {de Vaucouleurs}, A., {Corwin}, Jr., H.~G., {Buta},
  R.~J., {Paturel}, G., \& {Fouque}, P. 1991, {Third Reference Catalogue of
  Bright Galaxies} (Volume 1-3, XII, 2069 pp.~7 figs..~ Springer-Verlag Berlin
  Heidelberg New York)

\bibitem[{{Diemand} {et~al.}(2008){Diemand}, {Kuhlen}, {Madau}, {Zemp},
  {Moore}, {Potter}, \& {Stadel}}]{diema08}
{Diemand}, J., {Kuhlen}, M., {Madau}, P., {Zemp}, M., {Moore}, B., {Potter},
  D., \& {Stadel}, J. 2008, \nat, 454, 735

\bibitem[{{Dinescu} {et~al.}(2004){Dinescu}, {Keeney}, {Majewski}, \&
  {Girard}}]{dines04}
{Dinescu}, D.~I., {Keeney}, B.~A., {Majewski}, S.~R., \& {Girard}, T.~M. 2004,
  \aj, 128, 687

\bibitem[{{D'Onghia} \& {Lake}(2008)}]{dongh08}
{D'Onghia}, E. \& {Lake}, G. 2008, \apjl, 686, L61

\bibitem[{{Fellhauer} {et~al.}(2006){Fellhauer}, {Belokurov}, {Evans},
  {Wilkinson}, {Zucker}, {Gilmore}, {Irwin}, {Bramich}, {Vidrih}, {Wyse},
  {Beers}, \& {Brinkmann}}]{fellh06}
{Fellhauer}, M., {Belokurov}, V., {Evans}, N.~W., {Wilkinson}, M.~I., {Zucker},
  D.~B., {Gilmore}, G., {Irwin}, M.~J., {Bramich}, D.~M., {Vidrih}, S., {Wyse},
  R.~F.~G., {Beers}, T.~C., \& {Brinkmann}, J. 2006, \apj, 651, 167

\bibitem[{{Gilmore} {et~al.}(2007){Gilmore}, {Wilkinson}, {Wyse}, {Kleyna},
  {Koch}, {Evans}, \& {Grebel}}]{gilmo07}
{Gilmore}, G., {Wilkinson}, M.~I., {Wyse}, R.~F.~G., {Kleyna}, J.~T., {Koch},
  A., {Evans}, N.~W., \& {Grebel}, E.~K. 2007, \apj, 663, 948

\bibitem[{{Ibata} {et~al.}(1994){Ibata}, {Gilmore}, \& {Irwin}}]{ibata94}
{Ibata}, R.~A., {Gilmore}, G., \& {Irwin}, M.~J. 1994, \nat, 370, 194

\bibitem[{{Ibata} {et~al.}(1997){Ibata}, {Wyse}, {Gilmore}, {Irwin}, \&
  {Suntzeff}}]{ibata97}
{Ibata}, R.~A., {Wyse}, R.~F.~G., {Gilmore}, G., {Irwin}, M.~J., \& {Suntzeff},
  N.~B. 1997, \aj, 113, 634

\bibitem[{{Irwin} {et~al.}(1990){Irwin}, {Bunclark}, {Bridgeland}, \&
  {McMahon}}]{irwin90}
{Irwin}, M.~J., {Bunclark}, P.~S., {Bridgeland}, M.~T., \& {McMahon}, R.~G.
  1990, \mnras, 244, 16P

\bibitem[{{Kallivayalil} {et~al.}(2006{\natexlab{a}}){Kallivayalil}, {van der
  Marel}, \& {Alcock}}]{kalli06a}
{Kallivayalil}, N., {van der Marel}, R.~P., \& {Alcock}, C. 2006{\natexlab{a}},
  \apj, 652, 1213

\bibitem[{{Kallivayalil} {et~al.}(2006{\natexlab{b}}){Kallivayalil}, {van der
  Marel}, {Alcock}, {Axelrod}, {Cook}, {Drake}, \& {Geha}}]{kalli06b}
{Kallivayalil}, N., {van der Marel}, R.~P., {Alcock}, C., {Axelrod}, T.,
  {Cook}, K.~H., {Drake}, A.~J., \& {Geha}, M. 2006{\natexlab{b}}, \apj, 638,
  772

\bibitem[{{Karachentsev} {et~al.}(2004){Karachentsev}, {Karachentseva},
  {Huchtmeier}, \& {Makarov}}]{karac04}
{Karachentsev}, I.~D., {Karachentseva}, V.~E., {Huchtmeier}, W.~K., \&
  {Makarov}, D.~I. 2004, \aj, 127, 2031

\bibitem[{{Klypin} {et~al.}(1999){Klypin}, {Kravtsov}, {Valenzuela}, \&
  {Prada}}]{klypi99}
{Klypin}, A., {Kravtsov}, A.~V., {Valenzuela}, O., \& {Prada}, F. 1999, \apj,
  522, 82

\bibitem[{{Knebe} {et~al.}(2004){Knebe}, {Gill}, {Gibson}, {Lewis}, {Ibata}, \&
  {Dopita}}]{knebe04}
{Knebe}, A., {Gill}, S.~P.~D., {Gibson}, B.~K., {Lewis}, G.~F., {Ibata}, R.~A.,
  \& {Dopita}, M.~A. 2004, \apj, 603, 7

\bibitem[{{Kroupa} {et~al.}(2005){Kroupa}, {Theis}, \& {Boily}}]{kroup05}
{Kroupa}, P., {Theis}, C., \& {Boily}, C.~M. 2005, \aap, 431, 517

\bibitem[{{Kunkel}(1979)}]{kunke79}
{Kunkel}, W.~E. 1979, \apj, 228, 718

\bibitem[{{Kunkel} \& {Demers}(1976)}]{kunke76}
{Kunkel}, W.~E. \& {Demers}, S. 1976, in The Galaxy and the Local Group, 241--+

\bibitem[{{Law} {et~al.}(2005){Law}, {Johnston}, \& {Majewski}}]{law05}
{Law}, D.~R., {Johnston}, K.~V., \& {Majewski}, S.~R. 2005, \apj, 619, 807

\bibitem[{{Li} {et~al.}(2007){Li}, {Gu}, {Zhao}, {Huang}, \& {Luo}}]{li07}
{Li}, S.-P., {Gu}, Q.-S., {Zhao}, Y.-H., {Huang}, J.-S., \& {Luo}, X.-L. 2007,
  Chinese Journal of Astronomy and Astrophysics, 7, 764

\bibitem[{{Li} \& {Helmi}(2008)}]{li08}
{Li}, Y.-S. \& {Helmi}, A. 2008, \mnras, 307

\bibitem[{{Libeskind} {et~al.}(2007){Libeskind}, {Cole}, {Frenk}, {Okamoto}, \&
  {Jenkins}}]{libes07}
{Libeskind}, N.~I., {Cole}, S., {Frenk}, C.~S., {Okamoto}, T., \& {Jenkins}, A.
  2007, \mnras, 374, 16

\bibitem[{{Libeskind} {et~al.}(2005){Libeskind}, {Frenk}, {Cole}, {Helly},
  {Jenkins}, {Navarro}, \& {Power}}]{libes05}
{Libeskind}, N.~I., {Frenk}, C.~S., {Cole}, S., {Helly}, J.~C., {Jenkins}, A.,
  {Navarro}, J.~F., \& {Power}, C. 2005, \mnras, 363, 146

\bibitem[{{Lynden-Bell}(1976)}]{lynde76}
{Lynden-Bell}, D. 1976, \mnras, 174, 695

\bibitem[{{Lynden-Bell}(1982{\natexlab{a}})}]{lynde82}
---. 1982{\natexlab{a}}, The Observatory, 102, 202

\bibitem[{{Lynden-Bell}(1982{\natexlab{b}})}]{lynde82a}
---. 1982{\natexlab{b}}, The Observatory, 102, 7

\bibitem[{{Lynden-Bell}(1983)}]{lynde83}
{Lynden-Bell}, D. 1983, in IAU Symposium, Vol. 100, Internal Kinematics and
  Dynamics of Galaxies, ed. E.~{Athanassoula}, 89--91

\bibitem[{{Mayer} {et~al.}(2007){Mayer}, {Kazantzidis}, {Mastropietro}, \&
  {Wadsley}}]{mayer07}
{Mayer}, L., {Kazantzidis}, S., {Mastropietro}, C., \& {Wadsley}, J. 2007,
  \nat, 445, 738

\bibitem[{{Metz} {et~al.}(2007){Metz}, {Kroupa}, \& {Jerjen}}]{metz07}
{Metz}, M., {Kroupa}, P., \& {Jerjen}, H. 2007, \mnras, 374, 1125

\bibitem[{Metz {et~al.}(2009)Metz, Kroupa, \& Jerjen}]{metz09}
Metz, M., Kroupa, P., \& Jerjen, H. 2009, preprint arXiv:0901.1658

\bibitem[{{Metz} {et~al.}(2008){Metz}, {Kroupa}, \& {Libeskind}}]{metz08}
{Metz}, M., {Kroupa}, P., \& {Libeskind}, N.~I. 2008, \apj, 680, 287

\bibitem[{{Mirabel} {et~al.}(1992){Mirabel}, {Dottori}, \& {Lutz}}]{mirab92}
{Mirabel}, I.~F., {Dottori}, H., \& {Lutz}, D. 1992, \aap, 256, L19

\bibitem[{{Moore} {et~al.}(1999){Moore}, {Ghigna}, {Governato}, {Lake},
  {Quinn}, {Stadel}, \& {Tozzi}}]{moore99}
{Moore}, B., {Ghigna}, S., {Governato}, F., {Lake}, G., {Quinn}, T., {Stadel},
  J., \& {Tozzi}, P. 1999, \apjl, 524, L19

\bibitem[{{Naab} \& {Burkert}(2003)}]{naab03}
{Naab}, T. \& {Burkert}, A. 2003, \apj, 597, 893

\bibitem[{{Palma} {et~al.}(2002){Palma}, {Majewski}, \& {Johnston}}]{palma02}
{Palma}, C., {Majewski}, S.~R., \& {Johnston}, K.~V. 2002, \apj, 564, 736

\bibitem[{{Piatek} {et~al.}(2007){Piatek}, {Pryor}, {Bristow}, {Olszewski},
  {Harris}, {Mateo}, {Minniti}, \& {Tinney}}]{piate07}
{Piatek}, S., {Pryor}, C., {Bristow}, P., {Olszewski}, E.~W., {Harris}, H.~C.,
  {Mateo}, M., {Minniti}, D., \& {Tinney}, C.~G. 2007, \aj, 133, 818

\bibitem[{{Piatek} {et~al.}(2008){Piatek}, {Pryor}, \& {Olszewski}}]{piate08}
{Piatek}, S., {Pryor}, C., \& {Olszewski}, E.~W. 2008, \aj, 135, 1024

\bibitem[{{Piatek} {et~al.}(2003){Piatek}, {Pryor}, {Olszewski}, {Harris},
  {Mateo}, {et~al.}}]{piate03}
{Piatek}, S., {Pryor}, C., {Olszewski}, E.~W., {Harris}, H.~C., {Mateo}, M.,
  {et~al.} 2003, \aj, 126, 2346

\bibitem[{{Sales} {et~al.}(2007{\natexlab{a}}){Sales}, {Navarro}, {Abadi}, \&
  {Steinmetz}}]{sales07b}
{Sales}, L.~V., {Navarro}, J.~F., {Abadi}, M.~G., \& {Steinmetz}, M.
  2007{\natexlab{a}}, \mnras, 379, 1475

\bibitem[{{Sales} {et~al.}(2007{\natexlab{b}}){Sales}, {Navarro}, {Abadi}, \&
  {Steinmetz}}]{sales07a}
---. 2007{\natexlab{b}}, \mnras, 379, 1464

\bibitem[{{Salucci} \& {Burkert}(2000)}]{saluc00}
{Salucci}, P. \& {Burkert}, A. 2000, \apjl, 537, L9

\bibitem[{{Stoehr} {et~al.}(2002){Stoehr}, {White}, {Tormen}, \&
  {Springel}}]{stoeh02}
{Stoehr}, F., {White}, S.~D.~M., {Tormen}, G., \& {Springel}, V. 2002, \mnras,
  335, L84

\bibitem[{{Strigari} {et~al.}(2007){Strigari}, {Bullock}, {Kaplinghat},
  {Diemand}, {Kuhlen}, \& {Madau}}]{strig07}
{Strigari}, L.~E., {Bullock}, J.~S., {Kaplinghat}, M., {Diemand}, J., {Kuhlen},
  M., \& {Madau}, P. 2007, \apj, 669, 676

\bibitem[{{Tollerud} {et~al.}(2008){Tollerud}, {Bullock}, {Strigari}, \&
  {Willman}}]{tolle08}
{Tollerud}, E.~J., {Bullock}, J.~S., {Strigari}, L.~E., \& {Willman}, B. 2008,
  \apj, 688, 277

\bibitem[{{Tully} \& {Fisher}(1987)}]{tully87}
{Tully}, R.~B. \& {Fisher}, J.~R. 1987, {Atlas of Nearby Galaxies} (Annales de
  Geophysique)

\bibitem[{{Tully} {et~al.}(2006){Tully}, {Rizzi}, {Dolphin}, {Karachentsev},
  {Karachentseva}, {Makarov}, {Makarova}, {Sakai}, \& {Shaya}}]{tully06}
{Tully}, R.~B., {Rizzi}, L., {Dolphin}, A.~E., {Karachentsev}, I.~D.,
  {Karachentseva}, V.~E., {Makarov}, D.~I., {Makarova}, L., {Sakai}, S., \&
  {Shaya}, E.~J. 2006, \aj, 132, 729

\bibitem[{{Walsh} {et~al.}(1990){Walsh}, {van Gorkom}, {Bies}, {Katz}, {Knapp},
  \& {Wallington}}]{walsh90}
{Walsh}, D.~E.~P., {van Gorkom}, J.~H., {Bies}, W.~E., {Katz}, N., {Knapp},
  G.~R., \& {Wallington}, S. 1990, \apj, 352, 532

\bibitem[{{Walter} {et~al.}(2006){Walter}, {Martin}, \& {Ott}}]{walte06}
{Walter}, F., {Martin}, C.~L., \& {Ott}, J. 2006, \aj, 132, 2289

\bibitem[{{Weilbacher} {et~al.}(2003){Weilbacher}, {Duc}, \&
  {Fritze-v.~Alvensleben}}]{weilb03}
{Weilbacher}, P.~M., {Duc}, P.-A., \& {Fritze-v.~Alvensleben}, U. 2003, \aap,
  397, 545

\bibitem[{{Wu} {et~al.}(2008){Wu}, {Famaey}, {Gentile}, {Perets}, \&
  {Zhao}}]{wu08}
{Wu}, X., {Famaey}, B., {Gentile}, G., {Perets}, H., \& {Zhao}, H. 2008,
  \mnras, 386, 2199

\bibitem[{{Zhao}(1998)}]{zhao98}
{Zhao}, H. 1998, \apjl, 500, L149+

\bibitem[{Zwicky(1956)}]{zwick56}
Zwicky, F. 1956, Ergebnisse der exakten Naturwissenschaften, 29, 34

\end{thebibliography}

\clearpage

\clearpage

\clearpage

\end{document}